\documentclass[twocolumn, secnumarabic, amssymb, nobibnotes, aps]{revtex4-1}
\usepackage{amsmath,amstext,bm,appendix,graphicx,multirow,dcolumn,slashed}
\usepackage{color}
\usepackage{diagbox}
\usepackage{subfigure}
\usepackage{color}
\usepackage[bookmarks=false]{hyperref}
\hypersetup{colorlinks=true,citecolor=blue,linkcolor=blue,urlcolor=blue,pdfstartview=FitH,bookmarksopen=true}
\definecolor{green}{rgb}{0.8,0.98,0.83}
\pagecolor{white}
\date{\today}

\begin{document}
\title{Implementation and enhancement of nonreciprocal quantum synchronization with strong isolation in antiferromagnet-cavity systems}%
\author{Zhi-Bo Yang$^{1}$}%
\author{Hong-Yu Liu$^{1}$}%
\email{liuhongyu@ybu.edu.cn}
\author{Rong-Can Yang$^{2,3}$}%
\email{rcyang@fjnu.edu.cn}
\affiliation{$^{1}$Department of Physics, College of Science, Yanbian University, Yanji, Jilin 133002, China}
\affiliation{$^{2}$Fujian Provincial Key Laboratory of Quantum Manipulation and New Energy Materials, and College of Physics and Energy, Fujian Normal University, Fuzhou 350117, China}
\affiliation{$^{3}$Fujian Provincial Collaborative Innovation Center for Optoelectronic Semiconductors and Efficient Devices, Xiamen 361005, China}

\begin{abstract}
Sensitive signal detection and processing in classical world, especially in quantum regime, require nonreciprocal manipulation. In this paper we show how to achieve nonreciprocal quantum synchronization for two magnon modes in a two-sublattice antiferromagnet with strong isolation. The antiferromagnet is trapped in a cavity with two posts so that the two magnon modes not only couple to each other through a parametric-type interaction, but also interact with a same cavity respectively in a beam splitter-type and parametric-type ways. Under the condition of system's stability, we show that nonreciprocal quantum synchronization between two magnon modes is mainly dependent on resonance frequency of the cavity caused by direction of input currents. In addition, quantum synchronization is enhanced by the increase of interaction strength between two Bogoliubov modes and cavity mode. Moreover, numerical simulation results with parameters commonly used in current experiments show that the present scheme may be feasible.
\end{abstract}
\maketitle

$Introduction$.$-$Since Huygens first observed classical synchronization between two pendulum clocks (the isochronicity of the pendulum) in the 17th century~\cite{001}, there has been a great deal of research focused on it. 
Besides, with the continuous progress of science and technology, many researchers began to explore quantum synchronization which was first introduced by Mari and his co-workers~\cite{002}. 
They achieved this aim with cavity quantum electrodynamics~\cite{003,004}, atomic ensembles~\cite{005,006,007,008}, van der Pol oscillators~\cite{009,010,011,012,013}, Josephson junction~\cite{014,015}, optomechanics~\cite{016}, etc. 
But different from classical synchronization, strong nonlinear damping rates are desired to maintain steady states with low average populations in order to access quantum regime in their schemes.
Moreover, degree of quantum synchronization is usually very weak~\cite{017}.
To address this issue, some protocols are proposed with nonlinearity~\cite{018,019}, modulation manipulation ~\cite{020,021}, or adjustment of coupling rates between two subsystems~\cite{002,022,023,024}. 
However, the result is still not very satisfactory.
Recently, nonreciprocal devices attract much attention because they can enhance ability of quantum devices at the cost of PT symmetry.
Up to now, they have been utilized in several diverse fields such as chiral engineering~\cite{025,026,027,028,029}, invisible sensing~\cite{030,031}, backaction-immune information processing~\cite{032}, etc. 
In this paper, we will focus on nonreciprocal quantum synchronization (NQS) which may be applied for one-way magnon-magnon quantum synchronization, quantum diodes, etc.  

Additionally, cavity spintronics~\cite{033,034,035,036,037,038,039,040,041,042,043,044,045,046}, a new interdisciplinary subject, which studies a hybrid quantum system, seems to be a potential candidate for quantum information processing. The key element of the system is magnon, the quanta for a collective excitation of spins in a magnetic material, which can strongly interact with microwave photons via magnetic dipole interaction. Compared with other spin ensembles such as NV centers in diamond, spin density in a ferromagnet (e.g., yttrium iron garnet crystal, YIG for short) is much higher. In addition, magnon has low damping rate, and can be coupled to a variety of different systems. For example, magnon can interact with optical photons via magneto-optical effect\cite{046,047,048,049,050,051,052,053,054}, which then forms a new branch of discipline, i.e., cavity optomagnonics. Thus, a magnon-based hybrid system can provide a great opportunity to study entangled states~\cite{033,034,035}, squeezed states~\cite{036}, optical diodes~\cite{037}, one-way Gaussian steering~\cite{038}, etc. In this paper, we focus on cavity spintronics.

\begin{figure}
\centering
\includegraphics[width=1\linewidth,height=0.45\textheight]{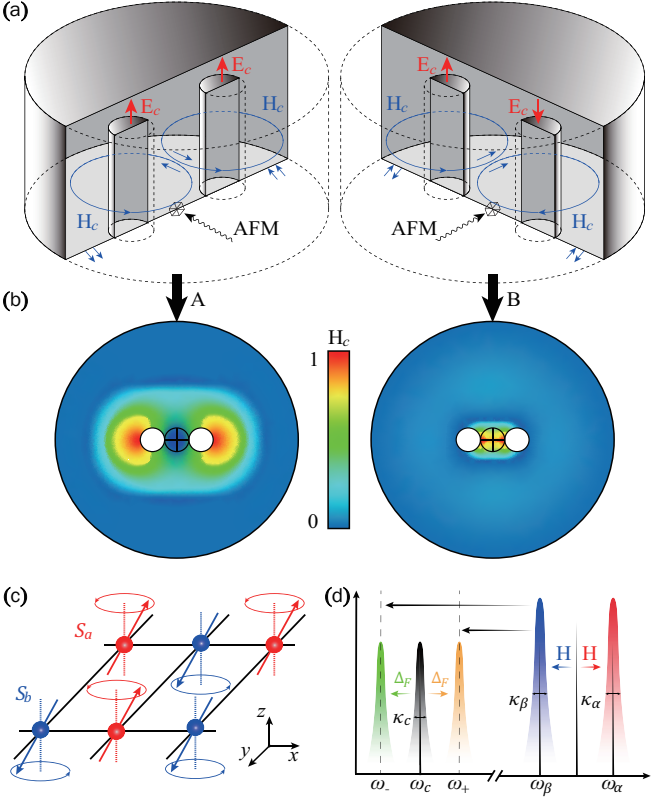}
\hspace{0in}%
\caption{(a) Sketch of the system. The two-sublattice AFM is placed near the reentrant-type cavity with two resonance posts. Cavity A demonstrates operation in the dark mode, and cavity B demonstrates operation in the bright mode. (b) The intensity and direction of local magnetic field distribution of two reentrant-type cavities (top view). (c) Schematic diagram of AFM crystal structure. (d) Frequency spectrum of the systems. The maximum value of quantum synchronization can be reached when the magnon mode $\beta$ (the superposition of magnon $a$ and magnon $b$) resonates with microwave signal. See text for more details.}
\label{fig1}
\end{figure}

It is all known that magnetic materials consist of ferromagnets (FOMs), ferrimagnets (FIMs) and antiferromagnets (AFMs). Compared with other two materials, AFMs have many characteristics including robustness against perturbation of external fields, negligible crosstalk with neighboring AFM elements because of absence of stray fields. Besides, distinctly different from a traditional cavity-spintronics system with FOMs or FIMs ~\cite{034} where beam splitter-type terms play a dominate role, both beam splitter-type and parametric-type coupling can simultaneously take effect with AFMs. Moreover, magnon-magnon coupling can be found in the frequency range of the order from several GHz to few THz with AFMs, generally much higher than that for FOMs or FIMs of the order of GHz~\cite{055,056}. For this reason, cavity spintronics with AFMs has attracted significant interest in the field of domain wall motion, skyrmions, magnetoresistance, magnetic switching, spin pumping, spin current transport, etc~\cite{057}. However, most of the works mainly focus on classical electrodynamics~\cite{039,058,059}. In order to better explore its application for NQS, we let AFMs confined in a reentrant cavity with two posts to form a hybrid system. The reentrant cavity is a closed (typically cylindrical) 3D microwave resonator which is built around a central post with an extremely small gap between one of the cavity walls and the post tip~\cite{060,062,063,064}, as shown in Fig.~\ref{fig1} (a). It leads the cavity field to possess spatial separation of the electric and magnetic components. However, the magnetic field distributed around each post has a fast radial decay in amplitude because most of the electric field is concentrated in the gap. It is noted that the cavity has two modes(i.e. dark and bright modes) due to the same or opposite direction of current flowing through the two posts. For the dark mode (Figure~\ref{fig1}, labeled A), the current direction in each post is the same, making the magnetic field  $\textbf{H}_{c}$ around each post curl in the same direction. And the magnetic field between them will effectively be canceled since the two magnetic directions are antiparallel to each other. Elsewhere, the field vectors are almost coaligned, thus enhancing the field. While for the bright mode (Figure~\ref{fig1}, labeled B), currents through two posts are always opposite, making $\textbf{H}_{c}$ for each post curl in the opposite direction. As a result, the total magnetic field $\textbf{H}_{c}$ is enhanced between posts and canceled elsewhere, as is demonstrated in Fig.~\ref{fig1} (b). Therefore, we can change resonance frequency of the cavity by the change of current directions flowing through two posts in the cavity.

$Model$ $and$ $equation$ $of$ $motion$.$-$ We assume that a two-sublattice AFM material where spins in sublattices $a$ and $b$ are oriented in opposite directions along the $z$-axis is confined in a reentrant cavity with two posts, as shown in Fig.~\ref{fig1} (c). 
Two posts are both fed into currents with the same (opposite) direction(s), leading the cavity to hold a dark (bright) mode. Noted that the magnetic component of each cavity mode lies perfectly in the plane perpendicular to the posts. Without loss of generality, we assume that the resonant frequency of the bright (dark) mode is $\omega_{+}$ ($\omega_{-}$), where $\omega_{\pm}=\omega_c\pm\Delta_F$ with $\omega_c$ and $\Delta_F$ being average frequency and drift frequency of cavity mode~\cite{060}. Thus, the Hamiltonian for the system is expressed by~\cite{040}
\begin{eqnarray}\label{e001}
\mathcal{H}&=&\mathcal{H}_{a+b}+\mathcal{H}_{c}+\mathcal{H}_{int},
\end{eqnarray}
where $\mathcal{H}_{a+b}$, $\mathcal{H}_{c}$, and $\mathcal{H}_{int}$ are, respectively, the Hamiltonian for spins in two sublattices $a$ and $b$, cavity and their interaction with the mathematical expression  
\begin{eqnarray}\label{e002}
\mathcal{H}_{a+b}&=&J\sum_{l,\delta}{(\textbf{S}_{a}^{l+\delta}\cdot \textbf{S}_{b}^{l}+\textbf{S}_{a}^{l}\cdot \textbf{S}_{b}^{l+\delta})}\nonumber\\
&&-\sum_{l}{g_a\mu(\textbf{H}_{an,a}+\textbf{H})\cdot \textbf{S}_{a}^{l}}\nonumber\\
&&-\sum_{l}{g_b\mu(\textbf{H}_{an,b}+\textbf{H})\cdot \textbf{S}_{b}^{l}},\nonumber\\
\mathcal{H}_{c}&=&\frac{1}{2}\int{(\epsilon_0 \textbf{E}_{c}^2+\mu_0 \textbf{H}_{c}^2)dxdydz},\nonumber\\
\mathcal{H}_{int}&=&-\sum_{l}{(g_a\mu \textbf{S}_{a}^l+g_{b}\mu \textbf{S}_{b}^l)\cdot \textbf{H}_{c}},
\end{eqnarray}
with
\begin{eqnarray}\label{e003}
\textbf{E}_{c}&=&i\sqrt{\frac{\omega_\pm}{4\epsilon_0V}}\sum_k {\left[\textbf{u}(\textbf{k})c_ke^{i\textbf{k}\cdot \textbf{q}}-\textbf{u}^*(\textbf{k})c_{k}^{\dag}e^{-i\textbf{k}\cdot \textbf{q}}\right]},\nonumber\\
\textbf{H}_{c}&=&i\sqrt{\frac{\mu_0\omega_\pm}{4V}}\sum_k {\textbf{k}\times \left[\textbf{u}(\textbf{k})c_ke^{i\textbf{k}\cdot \textbf{q}}-\textbf{u}^*(\textbf{k})c_{k}^{\dag}e^{-i\textbf{k}\cdot \textbf{q}}\right]}\nonumber\\
\end{eqnarray}
separately being electric and magnetic components. $c_{k}^{\dag}$ $(c_{k})$ represents creation (annihilation) operator of quantized electromagnetic modes in the cavity with wave vector $k$ and frequency $\omega_\pm$~\cite{039,040,041}. $\textbf{u}(k)$ is polarization vector, $V$ volume of the microcircular cavity, and $J$ ($>0$) exchange constant. $\textbf{S}_{o}^{l}$ ($o=a,b$) denotes spin operator at site $l$ in sublattice $o$, and $\delta$ represents displacement of two nearest spins. $\textbf{H}$ and $\textbf{H}_{an,o}$ are individually external static and anisotropy fields, respectively. Besides, $\mu_0$ is vacuum susceptibility, and $g_{o}$ and $\mu$ are separately effective $g$-factor for sublattice $o$ and Bohr magneton. 

Introducing Holstein-Primakoff transformation~\cite{039,040,041,065}
\begin{eqnarray}\label{e004}
S_{i}^{+,a}&=&\sqrt{2S_a}a_i,~~~~~~S_{i}^{+,b}=\sqrt{2S_b}b_i^{\dag},\nonumber\\
S_{i}^{-,a}&=&\sqrt{2S_a}a_i^{\dag},~~~~~~S_{i}^{-,b}=\sqrt{2S_b}b_i,\nonumber\\
S_{i}^{z,a}&=&S_a-a_{i}^{\dag}a_{i},~~~~~S_{i}^{z,b}=b_{i}^{\dag}b_{i}-S_b,
\end{eqnarray}
with $S_i^{\pm,o}=S_i^{x,o}\pm iS_{i}^{y,o}$ being the raising/lowering spin operator at site $i$ in sublattice $o$, and $a_{i}$, $b_{i}$ $(a_{i}^{\dag}, b_{i}^{\dag})$ corresponding annihilation (creation) operators for local excitation of spin ensemble at site $i$ in each sublattice, we can deduce an operator for the collective mode (i.e., magnon mode) $o_k$ in sublattice $o$ with $o_k=N^{-1/2}\sum_ie^{-i\vec{k}\cdot \vec{r}_i}o_i$. Then, the Hamiltonian (\ref{e002}) can be rewritten in phase space as
\begin{eqnarray}\label{e005}
\mathcal{H}&=&\mathcal{H}_{a+b}+\mathcal{H}_{c}+\mathcal{H}_{int}\nonumber\\
&=&\sum_k{\left[\omega_a a_{k}^{\dag}a_{k}+\omega_b b_{k}^{\dag} b_{k}+g_{ab} (a_{k}^{\dag} b_{k}^{\dag}+h.c.)\right]}\nonumber\\
&&+\sum_k{\left[\omega_{c} c_{k}^{\dag} c_{k}+g_{ac} (a_{k}^{\dag}c_{k}^{\dag}+h.c.)+g_{bc} (b_{k}^{\dag} c_{k}+h.c.)\right]},\nonumber\\
\end{eqnarray}
where $\omega_a=H_{ex,b}+H_{an,a}+H$ and $\omega_b=H_{ex,a}+H_{an,b}-H$ with $H_{ex,o}=2zJS_{o}^2$, $S_{o}$ magnitude of spin vector and $z$ being coordination number. $g_{ab}=\sqrt{H_{ex,a}H_{ex,b}}\cos(k\delta)$ denotes coupling rate between the two magnon modes, $g_{oc}=\sqrt{\mu_{0}\omega_{\pm}S_{o}N/2V}$ denotes the interaction strength between magnon $o$ and photon mode $c$ with $N$ the number of spins in each sublattice. In the long-wavelength limit ($k=0$)~\cite{039,040,041}, the Hamiltonian reads
\begin{eqnarray}\label{e006}
\mathcal{H}&=&\omega_{a} a^{\dag}a+\omega_b b^{\dag} b+g_{ab} (a^{\dag} b^{\dag}+h.c.)\nonumber\\
&&+\omega_{\pm} c^{\dag} c+g_{ac} (a^{\dag}c^{\dag}+h.c.)+g_{bc} (b^{\dag} c+h.c.).
\end{eqnarray}

After that, dissipative dynamics of the system can be described by a set of quantum Langevin equations (QLEs)~\cite{065}:
\begin{eqnarray}\label{e007}
\dot{a}&=&-(i\omega_a+\kappa_a)a-ig_{ab}b^{\dag}-ig_{ac}c^{\dag}+\sqrt{2\kappa_a}a^{in},\nonumber\\
\dot{b}&=&-(i\omega_b+\kappa_b)b-ig_{ab}a^{\dag}-ig_{bc}c+\sqrt{2\kappa_b}b^{in},\nonumber\\
\dot{c}&=&-(i\omega_{\pm}+\kappa_c)c-ig_{ac}a^{\dag}-ig_{bc}b+\sqrt{2\kappa_c}c^{in},
\end{eqnarray}
where $\kappa_{a,b}$ and $\kappa_c$ represent damping rates of two AFM magnon modes ($a$ and $b$) and cavity mode $c$, respectively. $a^{in}$, $b^{in}$ and $c^{in}$ individually stand for input noise operators for the corresponding modes $a$, $b$ and $c$, which are zero mean and characterized by the following correlation functions: $\langle o^{in}(t)o^{in\dagger}(t^{\prime})\rangle=\delta (t-t^{\prime})$ ($o=a,b,c$). If we further introduce quadrature components ($X_a$, $Y_a$, $X_b$, $Y_b$, $X_c$, $Y_c$) for two AFM magnon modes and cavity mode with $ X_{o}=(o+o^\dagger)/\sqrt{2}$, $ Y_{o}=(o-o^\dagger)/\sqrt{2}i$ ($o=a,b,c$), then we obtain
\begin{eqnarray}\label{e008}
\dot{u}(t)&=&\mathcal{A}u(t)+n(t),
\end{eqnarray}
where $u(t)=[X_{a}(t), Y_{a}(t), X_{b}(t), Y_{b}(t), X_{c}(t), Y_{c}(t)]^{T}$ and $ n(t)=[\sqrt{2\kappa_a}X_{a}^{in}(t),\sqrt{2\kappa_a}Y_{a}^{in} (t),\sqrt{2\kappa_b}X_{b}^{in}(t),\sqrt{2\kappa_b}\\Y_{b}^{in}(t),\sqrt{2\kappa_c}X_{c}^{in}(t),\sqrt{2\kappa_c}Y_{c}^{in}(t)]^T $ are, respectively, vectors for quantum fluctuations and noises, and drift matrix reads
\begin{eqnarray}\label{e009}
\mathcal{A}=
\begin{bmatrix} 
-\kappa_{a} & \omega_a & 0 & -g_{ab} & 0 & -g_{ac}\\ 
-\omega_a &-\kappa_{a}&-g_{ab}&0&-g_{ac}&0\\
0&-g_{ab}&-\kappa_{b}&\omega_b&0&g_{bc}\\
-g_{ab}&0&-\omega_b&-\kappa_{b}&-g_{bc}&0\\
0&-g_{ac}&0&g_{bc}&-\kappa_{c}&\omega_{\pm}\\
-g_{ac}&0&-g_{bc}&0&-\omega_{\pm}&-\kappa_{c}\\
\end{bmatrix}.
\end{eqnarray}

Since we are using linearized QLEs, Gaussian nature of input states will be preserved during time evolution for the system. Thus, quantum fluctuations will remain in a continuous three-mode Gaussian state which can be completely characterized by a $6\times6$ covariance matrix (CM) $\mathcal{V}$ with  $ \mathcal{V}_{ij}(t,t^{\prime})=\langle u_i(t)u_j(t^{\prime})+u_j(t^{\prime})u_i(t)\rangle/2$, ($i$, $j=1$, $2$, $...$, $6$). In addition, we only consider the system to be in the steady state when $t\to +\infty$. Then the vector can be obtained straightforwardly by solving the steady Lyapunov equation
\begin{eqnarray}\label{e010}
\mathcal{A}\mathcal{V}+\mathcal{V}\mathcal{A}^{T}&=&-\mathcal{D}
\end{eqnarray}
with $ \mathcal{D}=$ diag $[\kappa_a,\kappa_a,\kappa_b,\kappa_b,\kappa_c,\kappa_c] $ define through $ \mathcal{D}_{ij}\delta (t-t^{\prime})=\langle n_i(t)n_j(t^{\prime})+n_j(t^{\prime})n_i(t) \rangle/2$. 

$Nonreciprocal$ $quantum$ $synchronization$.$-$Similar to the definition proposed by  Mari and his co-workers~\cite{002}, we define the relative measure of quantum synchronization for two magnon modes as
\begin{eqnarray}\label{e011}
\mathcal{S}&=&\langle X_{-}(t)^2+Y_{-}(t)^2 \rangle ^{-1},
\end{eqnarray}
with $X_{-}(t)=[X_a(t)-X_b(t)]/\sqrt{2}$ and $Y_{-}(t)=[Y_a(t)-Y_b(t)]/\sqrt{2}$. According to Eq.~(\ref{e011}) and Heisenberg noncommutation relation$\langle X_{-}(t)^2 \rangle\langle Y_{-}(t)^2 \rangle\geq\frac{1}{4}$, we have $\mathcal{S}\leq1$. It is worth noting that the higher $\mathcal{S}$ is, the more synchronous the two magnon modes are, and $\mathcal{S}=1$ means they are completely synchronized. Due to the input current direction is different, frequencies of bright mode and dark mode are different, leading to two different situations of quantum synchronization for the two AFM magnon modes. In other words, quantum synchronization of the two AFM magnon modes for bright and dark modes are nonreciprocal. Hence, the degree of quantum synchronization $\mathcal{S}$ is divided to two different values $\mathcal{S}_{12}$ for the case of bright mode and $\mathcal{S}_{21}$ for the other case of dark mode. To obtain a more comprehensive understanding of the NQS, we use synchronization isolation ratio (SIR) to determine the different between $\mathcal{S}_{12}$ and $\mathcal{S}_{21}$, whose mathematical form is expressed as~\cite{066}
\begin{eqnarray}\label{e012}
\mathcal{S}_{iso}&=&20\times\log_{10}\zeta
\end{eqnarray}
with $\zeta=\max(\arrowvert \mathcal{S}_{12}/\mathcal{S}_{21}\arrowvert,\arrowvert \mathcal{S}_{21}/\mathcal{S}_{12}\arrowvert)$. The greater $\mathcal{S}_{iso}$ is, the higher the SIR degree is. 
\begin{figure}
\centering
\includegraphics[width=0.96\linewidth,height=0.35\textheight]{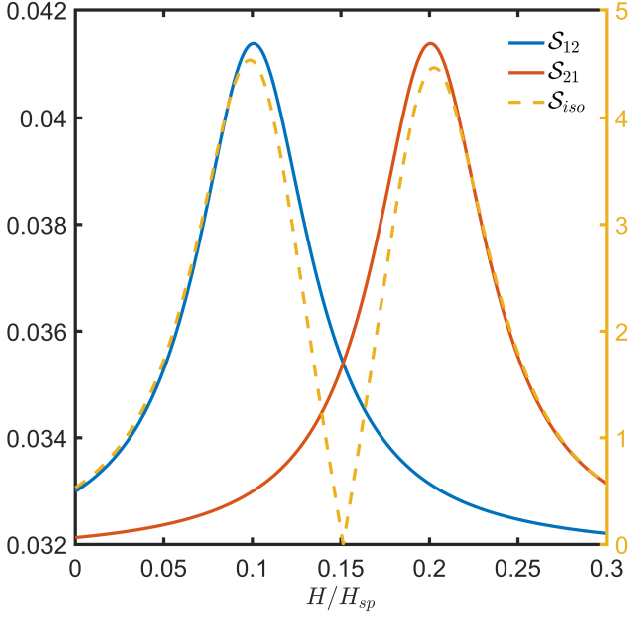}
\hspace{0in}%
\caption{Two-sublattice AFM NQS $\mathcal{S}_{12}$ ($\omega_c=\omega_{+}$, bright mode), $\mathcal{S}_{21}$ ($\omega_c=\omega_{-}$, dark mode), and synchronization isolation ratio versus the external static field $H$. We take $\omega_c/H_{sp}=0.85$, $\Delta_{F}/H_{sp}=0.05$, and see text for the details of the other parameters.}
\label{fig2}
\end{figure}

To show NQS more clearly, we draw curves for the degree of quantum synchronization $\mathcal{S}_{12}$, $\mathcal{S}_{21}$ and SIR $\mathcal{S}_{iso}$ as a function of external static field $H$ in Fig.~\ref{fig2}, where we have chosen experimental parameters~\cite{039,040,041}: $S_a=S_b=S$ because of the sublattice permutation symmetry of AFMs~\cite{039,040,041}, $H_{an}= 0.0163H_{ex}$, $H_{sp}=\sqrt{H_{an}(H_{an}+2H_{ex})} $ which is called the spin-flop field, $\omega_c/H_{sp}=0.85$, $g_{ac}=g_{bc}=g= 0.01H_{ex}$, $g_{ab}/H_{ex}=1$, $\kappa_{a}=\kappa_{b}=\kappa$, $\kappa_c=3\kappa_a=3\kappa_b=0.003H_{ex}$, and $\Delta_F/H_{sp}=0.05$. It is noted that all parameters chosen above satisfy the requirement of steady conditions, which is guaranteed by the negative real component of eigenvalues of the drift matrix $ \mathcal{A} $. Based on Fig.~\ref{fig2}, it can be seen that the function of $\mathcal{S}_{12}$ and $\mathcal{S}_{21}$ with the external static field $H$ is distinctly different, and the maximum value of $\mathcal{S}_{12}$ ($\mathcal{S}_{21}$) appears at $H/H_{sp}=0.1$ (0.2) where $S_{iso}$ nearly reaches the maximum value. This phenomena indicates that quantum synchronization of the two AFM magnon modes for the cavity being in bright and dark modes is nonreciprocal. In addition, Fig.~\ref{fig2} also tells us that $\mathcal{S}_{iso}=0$ (i.e., $\mathcal{S}_{12}=\mathcal{S}_{21}$) with the choice of $ H/H_{sp}=0.15$, meaning that the system reaches the impedance-matching condition~\cite{067}, and at the same time the two AFM magnon modes are completely reciprocal.

\begin{figure}
	\centering
	\includegraphics[width=0.96\linewidth,height=0.35\textheight]{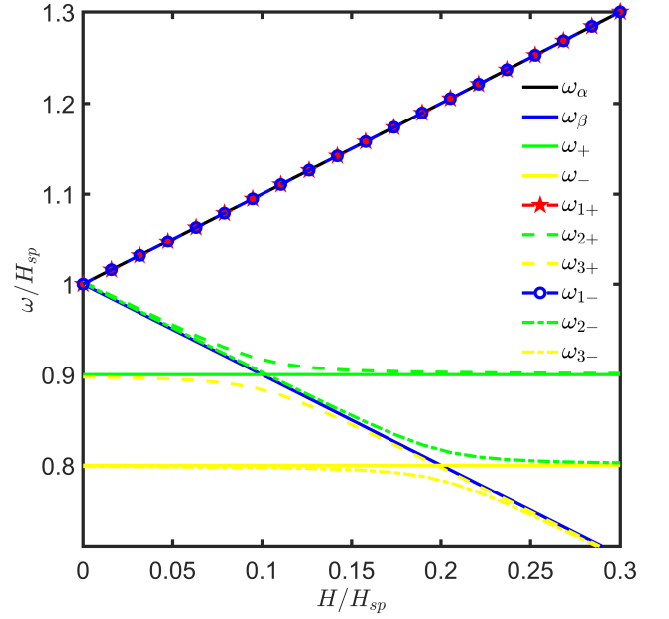}
	\hspace{0in}%
	\caption{The dispersion relations for an AFM. Obviously, according to current input direction, two anticrossing gap appear at $H=0.1H_{sp}$ and $H=0.2H_{sp}$ respectively. The parameters are the same as those used in Fig.~\ref{fig2}.}
	\label{fig3}
\end{figure}
In order to better understand NQS of the two AFM magnon modes, we diagonalize the first three terms of $\mathcal{H}$ described in Eq.~(\ref{e006}) by introducing two orthogonal Bogoliubov modes~\cite{039,040,041,059}: $\alpha=\cosh\theta a-\sinh \theta b^{\dag} $ and $\beta=\cosh\theta b-\sinh\theta a^{\dag} $ with $\tanh{2\theta}=-2g_{ab}/(\omega_a+\omega_b)$. After that, the Hamiltonian $\mathcal{H}$ can be rewritten as
\begin{eqnarray}\label{e013}
\mathcal{H}&=&\omega_{\alpha}\alpha^{\dag}\alpha+\omega_{\beta}\beta^{\dag}\beta+\omega_{\pm}c^{\dag}c\nonumber\\
&&+g_{\alpha c}(\alpha^{\dag}c^{\dag}+h.c.)+g_{\beta c}(\beta c^{\dag}+h.c.)\nonumber\\
&=&\Psi^{\dagger}\left(\begin{matrix} \omega_{\alpha}&0&g_{\alpha c}\\0&\omega_{\beta}&g_{\beta c}\\g_{\alpha c}&g_{\beta c}&\omega_{\pm}
\end{matrix}\right)\Psi
\end{eqnarray}
with $\Psi = (\alpha,\beta^{\dagger},c^{\dagger})^{\dagger}$, where $g_{\alpha c}=g_{ac}\cosh \theta+g_{bc}\sinh \theta$, $g_{\beta c}=g_{ac} \sinh \theta+g_{bc}\cosh \theta$, $ \omega_{\alpha}=H_{sp}+H $, and $\omega_{\beta}=H_{sp}-H$. Then, we depict dispersion relation of the Hamiltonian $\mathcal{H}$ in Fig.~\ref{fig3} with same parameters chosen in Fig.~\ref{fig2}, where eigenvalues of the system ($\omega_{1\pm}$, $\omega_{2\pm}$, $\omega_{3\pm}$) can be obtained by the diagonalization of the matrix in Eq.~(\ref{e013}). For comparison, we also describe frequencies of the two Bogoliubov modes and cavity mode ($\omega_{\alpha/\beta/\pm}$) as the function of $H$ in the same figure. We see that anticrossing gap appear at $ H/H_{sp}=0.1$, which meaning that the cavity mode is resonant with the Bogoliubov mode $\beta$, i.e., $\omega_{\beta}=$ $\omega_c$ appears at $H/H_{sp}=0.1$ (0.2) for $\omega_c=\omega_{+}$ ($\omega_{-}$), meaning that the cavity mode is resonant with the Bogoliubov mode $\beta$, i.e., $\omega_{\beta}=\omega_{+}$ ($\omega_{-}$) appears at $H/H_{sp}=0.1$ (0.2). At this time, eigenstates with frequencies $\omega_{2+}$, $\omega_{3+}$ (for $\omega_c=\omega_{+}$) and $\omega_{2-}$, $\omega_{3-}$ (for $\omega_c=\omega_{-}$) of the coupled system corresponds to the superposition state of the two magnon modes and cavity mode [see Fig.~\ref{fig1} (d)]. While the eigenstate with frequency $\omega_{1\pm}=\omega_{\alpha}$ is a dark mode~\cite{039,040}. Therefore, the essence of NQS is that the existence of cavity mode causes the two magnon modes to be further squeezed. 

\begin{figure}
	\centering
	\includegraphics[width=0.96\linewidth,height=0.35\textheight]{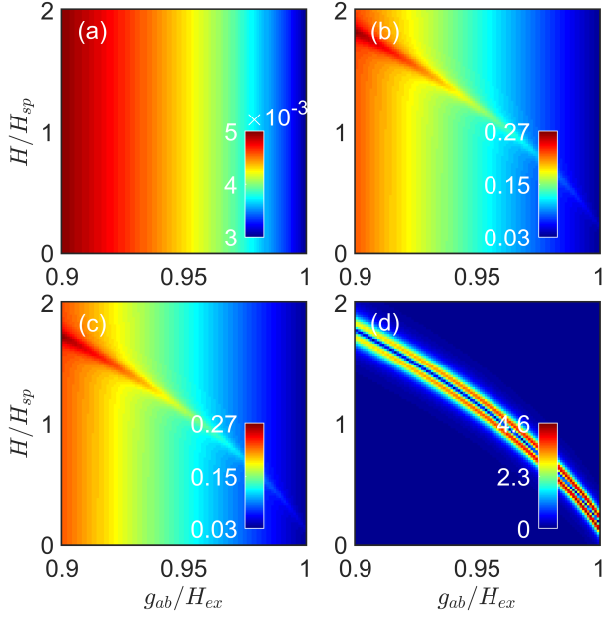}
	\hspace{0in}%
	\caption{(a) Effective coupling rate ($g_{\alpha c}=g_{\beta c}$) between Bogoliubov modes and cavity photons, (b) degree of quantum synchronization between two magnon modes $S_{12}$, (c) $S_{21}$, and (d) SIR $S_{iso}$ versus coupling rate between magnons $g_{ab}$ and external static field $H$. All parameters chosen are the same as those used in Fig.~\ref{fig2}.}
	\label{fig4}
\end{figure}
According to numerical simulation results described in above passages, it shoud be noted that the obtained quantum synchronization between two magnon modes is very low. In order to solve this problem, we analyze the influence of external magnetic field $H$ and $g_{ab}$ in Fig.~\ref{fig4} with the other parameters being the same with that used in Fig.~\ref{fig2}. Based on Fig.~\ref{fig4} (a), we find that interacting strength between two Bogoliubov modes and cavity mode ($g_{\alpha c}$ and $g_{\beta c}$) is almost inversely proportion to the coupling rate $g_{ab}$ between two sublattices of AFMs, but scarcely sensitive to the external magnetic field $H$. Similarly, from Fig.~\ref{fig4} (b) and (c), it is clearly shown that no matter $S_{12}$ for bright mode or $S_{21}$ for dark mode will gradually increase with the decrease of $g_{ab}$, and both of them are hardly affected by $H$. However, there is a special region where $S_{12}$ ($S_{21}$) can be strengthened more largely with the decrease of $g_{ab}$ and the increase $H$ at the same time, meaning that NQS can be realized. For example, $S_{12}\simeq0.27$ and $S_{21}\simeq0.24$ when $g_{ab}=0.9H_{ex}$ and $H=1.8H_{sp}$. In order to show it more intuitionally,  we draw Fig.~\ref{fig4} (d) for $S_{iso}$ with the function of $g_{ab}$ and $H$. It is clearly illustrated from the subgraph that maximum $S_{iso}$ is almost not reduced with the increase of $S_{12/21}$ when we adjust $g_{ab}$ and $H$ to appropriate values at the same time. Thus, we can enhance NQS with strong isolation.

\begin{table*}
	\caption{\label{tab1}%
		List of parameter values used in the simulation in Fig.~\ref{fig5}. The exchange field of DPPH is estimated from its N$\acute{e}$el temperature. The damping rate of $ {\rm MnF_2} $ is estimated from the linewidth measured in the experiments~\cite{069}.}
	\begin{ruledtabular}
		\begin{tabular}{cccccc}
			\multicolumn{1}{c}{\textrm{Material}}&
			\multicolumn{1}{c}{\textrm{$H_{ex}$(T)}}&
			\multicolumn{1}{c}{\textrm{$H_{an}/H_{ex}$}}&
			\multicolumn{1}{c}{\textrm{$g/H_{ex}$}}&
			\multicolumn{1}{c}{\textrm{$\kappa/H_{ex}$}}&
			\multicolumn{1}{c}{\textrm{$\kappa_c/H_{ex}$}}\\
			\hline
			\mbox{DPPH}~\cite{068}&\mbox{1.73}&\mbox{$ 1.8\times10^{-2} $}&\mbox{$ 8\times10^{-4} $}&\mbox{$ 1.05\times10^{-5} $}&$ 6.12\times10^{-4} $\\
			\mbox{$ {\rm MnF_2} $}~\cite{069}&51.5&$ 1.63\times10^{-2} $&$ 1\times10^{-3} $&$ 9.7\times10^{-6} $&$ 6\times10^{-4} $ \\
			\mbox{$ {\rm NaNiO_2} $}~\cite{070}&4.8&$ 7.3\times10^{-2} $&$ 1.2\times10^{-2} $&$ 1\times10^{-3} $&$ 5\times10^{-3} $ \\
			\mbox{$ {\rm NiO} $}~\cite{071}&524&$ 2.8\times10^{-3} $&$ 3\times10^{-4} $&$ 5\times10^{-4} $&$ 1\times10^{-4} $ \\
		\end{tabular}
	\end{ruledtabular}
\end{table*}

\begin{figure}
	\centering
	\includegraphics[width=0.96\linewidth,height=0.35\textheight]{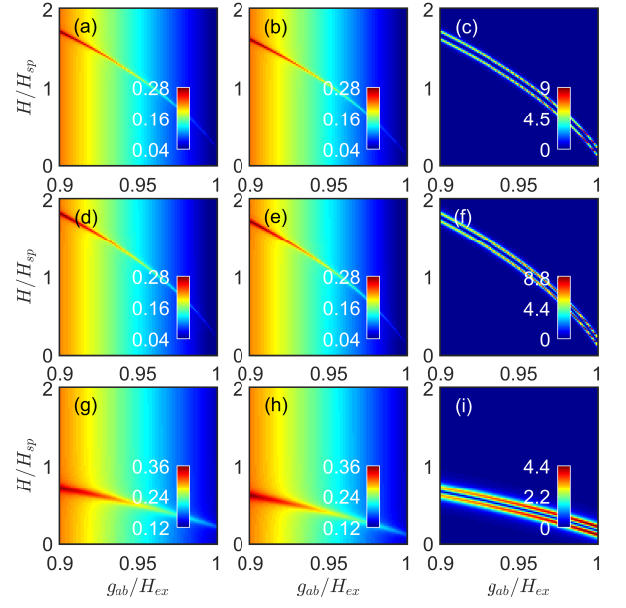}
	\hspace{0in}%
	\caption{The degree of quantum synchronization (a) (d) (g) $\mathcal{S}_{12}$, (b) (e) (h) $\mathcal{S}_{21}$ and (g) (h) (i) the SIR $\mathcal{S}_{iso}$ versus the coupling of between magnons $g_{ab}$ and external field $H$ for (a)--(c) DPPH~\cite{068}, (d)--(f) ${\rm MnF_2}$~\cite{069}, and (g)--(i) ${\rm NaNiO_2}$~\cite{070}. Please see the experimental parameters in TABLE~\ref{tab1}.}
	\label{fig5}
\end{figure}
In order to show universality of the present scheme, we select some AFMs to verify our theory, such as di(phenyl)-(2,4,6-trinitrophenyl)iminoazanium (${\rm DPPH}$)~\cite{068}, ${\rm MnF_2}$~\cite{069}, $ {\rm NaNiO_2}$~\cite{070}, and ${\rm NiO}$~\cite{071}, which is similar to that in Ref.~\cite{040}. Numerical simulation for degree of quantum synchronization $S_{12}$ and $S_{21}$ and SIR $S_{iso}$ for these materials is described in Fig.~\ref{fig5}, where exchange and anisotropy fields depend on magnetic parameters such as $H_{ex}=2zJS^2$ and $H_{an}=2KS^2$ with $K$ being the anisotropy coefficient~\cite{040}. In addition, they can experimentally be determined through magnetic resonance techniques~\cite{068,069,070}. For the sake of simplicity,  in this article we have unified the dimensions of the parameters as frequency units. For example, the resonance frequency of ${\rm NiO}$ at room temperature is about $\omega\simeq H_{ex}+H_{an}\simeq2\pi\times1.1$ THz, where $H_{ex}$ and $H_{an}$ contain the gyromagnetic ratio. Since $H_{an}/H_{ex}=zJ/K=0.0028\ll1$, we have $\omega/H_{ex}\simeq 1$. Moreover, the damping rate of ${\rm NiO}$ at room temperature is $\kappa/2\pi\simeq0.55$ GHz, i.e., $\kappa/H_{ex}\simeq5\times10^{-4}$ in Ref.~\cite{040,071}. For another examples, resonance frequency and damping rate of ${\rm NaNiO_2}$ are separately $2\pi\times0.84$ THz and $2\pi\times0.84$ GHz when environment temperature is $T=4$ K~\cite{070}. In this case, the cavity mode with resonant frequency $\omega_c/2\pi\simeq0.7$ THz (corresponds to $\omega_c/H_{sp}=0.85$ with the spin-flop field $H_{sp}=1.8$ T) should be used to match the frequency of magnon modes. Without loss of generality, remaining parameters are chosen as the same with Fig.~\ref{fig4}. On the basis of the graphs, we find that the implementation of magnon-magnon NQS with strong isolation for the other three AFMs is still valid.

$Feasibility$ $and$ $conclusion$.$-$In the last few paragraphs, let us focus on feasibility of the scheme and make a conclusion. The experimental setup is sketched in Fig.~\ref{fig1} (a). The key component is a two-sublattice AFM material~\cite{040,041} and a reentrant cavity~\cite{060,062,063,064}, which is a closed (typically cylindrical) 3D microwave resonator built around a central post with an extremely small gap between one of cavity walls and post tips. In fact, a reentrant cavity can be considered the 3D realization of a lumped LC circuit and thus can operate in a subwavelength regime. Microwave reentrant cavities as described have a wellconfined electrical field, but the magnetic field is distributed over quite some volume, which is significantly larger than that of a typical AFM. Thus, it is a necessity to focus this field into a relatively small region of the cavity, a problem that is solved by employing a double-post structure [see Fig.~\ref{fig1} (a)] which is the simplest case of a 3D reentrant cavity lattice~\cite{072}.

In the experiment~\cite{060}, for actual cavity dimensions (an internal cavity radius of 5 mm, cavity height 1.4 mm, post radius of 0.4 mm, and post gap of 73 $\mu$m, distance between the posts 1.5 mm), predicted resonance frequencies for dark and bright modes are 13.75 and 20.6 GHz, respectively. However, in our scheme, a cavity with a higher frequency should be used, which can be achieved by increasing the post gap size. However, our scheme is only a theoretical prediction, and in actual operation, the upper limit frequency of cavity field may not reach 0.7 THz. Therefore, a bias field with a larger amplitude is needed to adjust frequency of acoustic magnon to resonate with cavity mode with a lower frequency. For example, when the amplitude is chosen to be 1.1 T, cavity mode and acoustic magnon are resonant at 21 GHz, where the magnetic material ${\rm NaNiO_2}$ is used~\cite{073}. In addition, the distribution of magnetic field for two type of cavity modes are shown in Fig.~\ref{fig1} (b). The system described is similar to magnetic field of a current loop. In the plane of the loop, fields generated by any two opposite loop sections are added in phase within the bounded area of the loop, and out of phase outside the loop area. This strongly enhances the total field inside the loop and effectively cancels it outside. Thus, the cavity presented here can be understood as a two-dimensional current dipole. This field-focusing effect results in very high spatial overlap between the photon mode of the cavity and the magnon modes of the AFM, and thus the strong coupling between them.


In summary, we propose to achieve NQS in the system of a microwave reentrant cavity coupled to AFMs with a considerable SIR and flexible controllability by introducing two types of cavity modes (i.e., bright mode and dark mode). In order for the enhancement of NQS, the direct influence of external magnetic field $H$ and coupling strength between magnons $g_{ab}$ on quantum synchronization is demonstrated with some experimental parameters. We hope that this work may show some excellent applications in the field of quantum information processing that act either as a quantum transducer or a quantum memory, and provide some ideas for the realization and of nonreciprocal quantum correlations in other systems.

$Acknowledgments$.$-$This work is supported by Outstanding Young Talent Fund Project of Jilin Province (Grant No. 20180520223JH) and Fujian Natural Science Foundation (Grant Nos. 2018J01661 and 2019J01431).

\end{document}